\documentclass{sigchi}
\toappearbox{\large Accepted for publication at ACM IUI 2018.}

\usepackage{balance}       % to better equalize the last page
\usepackage{graphics}      % for EPS, load graphicx instead 
\usepackage[T1]{fontenc}   % for umlauts and other diaeresis
\usepackage{txfonts}
\usepackage{mathptmx}
\usepackage[pdflang={en-US},pdftex]{hyperref}
\usepackage{color}
\usepackage{booktabs}
\usepackage{textcomp}
\usepackage{bm}
\usepackage{changes}
\usepackage{makecell}
\usepackage{xspace}

\usepackage{microtype}        % Improved Tracking and Kerning
\usepackage{ccicons}          % Cite your images correctly!

\def\plaintitle{Detecting Low Rapport During Natural Interactions in Small Groups from Non-Verbal Behaviour}
\def\plainauthor{Philipp M\"uller, Michael Xuelin Huang, Andreas Bulling}

\def\plainkeywords{Social Signal Processing; Affective Computing; Facial Expressions; Body Posture; Speech Prosody; Personality Traits; Leadership; Dominance}

\newcommand{\change}[1]{#1}

\makeatletter
\def\url@leostyle{%
  \@ifundefined{selectfont}{
    \def\UrlFont{\sf}
  }{
    \def\UrlFont{\small\bf\ttfamily}
  }}
\makeatother
\def\pprw{8.5in}
\def\pprh{11in}

\definecolor{linkColor}{RGB}{6,125,233}
\hypersetup{%
  pdftitle={\plaintitle},
  pdfauthor={\plainauthor},
  pdfkeywords={\plainkeywords},
  pdfdisplaydoctitle=true, % For Accessibility
  bookmarksnumbered,
  pdfstartview={FitH},
  colorlinks,
  citecolor=black,
  filecolor=black,
  linkcolor=black,
  urlcolor=linkColor,
  breaklinks=true,
  hypertexnames=false
}

\begin{document}

\title{\plaintitle}

\numberofauthors{3}
\author{
  \alignauthor{Philipp M\"uller\\
    \affaddr{Max Planck Institute for Informatics, Saarland Informatics Campus}\\
    \email{\href{mailto:pmueller@mpi-inf.mpg.de}{\nolinkurl{pmueller@mpi-inf.mpg.de}}}}\\
  \alignauthor{Michael Xuelin Huang\\
    \affaddr{Max Planck Institute for Informatics, Saarland Informatics Campus}\\
    \email{\href{mailto:mhuang@mpi-inf.mpg.de}{\nolinkurl{mhuang@mpi-inf.mpg.de}}}}\\
  \alignauthor{Andreas Bulling\\
    \affaddr{Max Planck Institute for Informatics, Saarland Informatics Campus}\\
    \email{\href{mailto:bulling@mpi-inf.mpg.de}{\nolinkurl{bulling@mpi-inf.mpg.de}}}}\\
}

\maketitle

\begin{abstract}

%!TEX root = main.tex

Rapport, the close and harmonious relationship in which interaction partners are ``in sync'' with each other, was shown to result in smoother social interactions, improved collaboration, and improved interpersonal outcomes.
In this work, we are first to investigate automatic prediction of low rapport during natural interactions within small groups.
This task is challenging given that rapport only manifests in subtle non-verbal signals that are, in addition, subject to influences of group dynamics as well as inter-personal idiosyncrasies.
We record videos of unscripted discussions of three to four people using a multi-view camera system and microphones.
We analyse a rich set of non-verbal signals for rapport detection, namely facial expressions, hand motion, gaze, speaker turns, and speech prosody.
Using facial features, we can detect low rapport with an average precision of 0.7 (chance level at 0.25), while incorporating prior knowledge of participants' personalities can even achieve early prediction without a drop in performance.
We further provide a detailed analysis of different feature sets and the amount of information contained in different temporal segments of the interactions.

\end{abstract}

\category{H.5.m.}{Information Interfaces and Presentation
  (e.g. HCI)}{Miscellaneous}

\keywords{\plainkeywords}

%!TEX root = main.tex
\section{Introduction}
\iffalse
Summarize positive effects of rapport here.
\begin{itemize}
    \item Rapport predicts peer tutoring performance \cite{sinhacognitive,sinha2015fine}
    \item many examples of positive effects in introduction of \cite{gratch2007creating}
    \item here are also some examples right at beginngin of paper: \cite{zhao2014towards}
\end{itemize}

Contributions:
\begin{itemize}
\item New dataset, first dataset with rapport annotation in groups.
\item Prediction and analysis using Facial Action Units in groups for emergent leadership prediction.\michael{related?}
\item Prediction of rapport in groups.
\item Dyadic vs. indiv predictions.
\end{itemize}
\fi

\begin{figure}
  \centering
  \includegraphics[width=0.9\columnwidth]{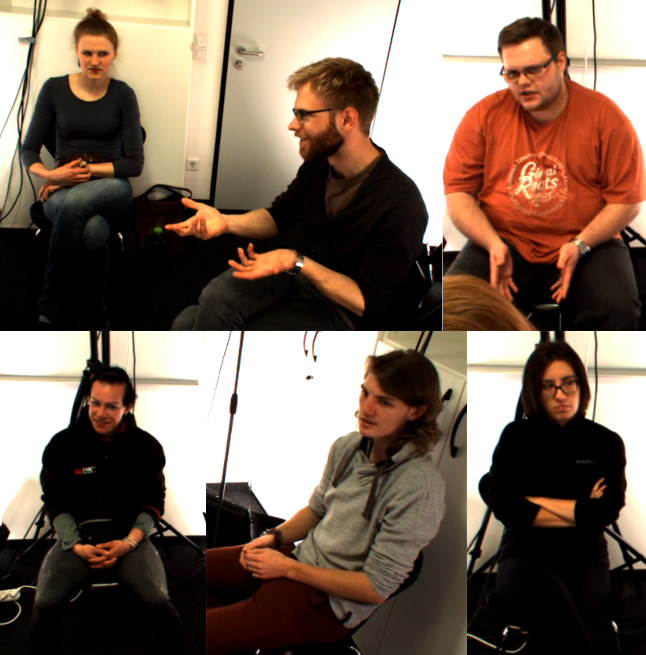}
  \caption{Example images of natural behaviours from the dataset.}~\label{teaser}
\end{figure}

Inter-personal conflicts are pervasive and can happen in a variety of social settings, from festivals, to family gatherings, to a bar or the classroom.
Many of these conflicts are the result of low rapport between interaction or conversation partners, or more specifically, the failure of a person to establish good rapport.
While a precise definition is difficult, rapport refers to the close and harmonious relationship in which interaction partners are ``in sync'' and can interact naturally and smoothly with each other.
Failure to build rapport can lead to mutual feelings of disharmony or, in the worst case, even verbal or physical hostility.
The fundamental importance of rapport for social interactions underlines the significant potential of developing intelligent user interfaces that are able to detect low rapport and to reduce or even avoid inter-personal conflicts.

While several previous works in social signal processing and affective computing investigated automatic detection of rapport during \textit{dyadic} (person-person or person-machine) interactions from verbal and non-verbal behaviour~\cite{cerekovic2016rapport,hagad2011predicting,wang2009rapport,zhao2016socially}, few works studied the link between non-verbal behaviour and rapport in larger groups \cite{lafrance1976group}.
No attempt has so far been made to automatically detect low rapport in \textit{multi-person} interaction settings.
This is despite the fact that much of our social life takes place in groups larger than two people, e.g. in business meetings or friend gatherings.
Detecting low rapport and performing an early intervention to avoid social conflicts can therefore have a significant and practical impact.

We present the first study on detecting the failure to establish rapport in natural multi-person interactions with a small number of people.
Given that no dataset exists that comprises small group interactions as well as annotations of felt rapport, we record a new dataset (see Figure \ref{teaser} for example images).
Based on this dataset, we then develop a multimodal approach to automatically detect low rapport from non-verbal behaviour.
Our approach is based on state-of-the-art methods to analyse facial expression and posture, as well as speech activities and prosodic features.
We further propose new features that exploit the mirroring effect by accounting for behaviour synchronisation among group members as well as cross-modal features that delineate simultaneous actions from different modalities.
Our results show that while facial features perform best when the full interaction is observed, prior information about participants' personalities can boost facial features to achieve the same performance while observing only the first third of the whole interaction.

The specific contributions of our work are three-fold:
1) We collect the first dataset for small group interactions with informative audio-visual signals and rich annotations, including \textit{felt rapport}, perceived leadership, dominance, competence and liking of the dyads in the group.
2) We propose a multimodal approach to low rapport detection that exploits both dyadic audiovisual information, such as facial action units and speech prosody, as well as group information, such as cross-modal features and mirroring effects, and potential prior knowledge of the participants' personalities.
3) We provide an in-depth performance evaluation of our method and identify key features and time segments that are most important for rapport detection in this setting.

%!TEX root = main.tex

\section{Related Work}

We first summarise prior works that aimed to predict social aspects in multi-person interactions.
We then focus on rapport as a particularly important social aspect, followed by computational methods to predict rapport in dyadic interactions.

\subsection{Automatic Analysis of Multi-Person Interactions}

While a major part of research in social signal processing and affective computing focuses on the analysis of dyadic interactions~\cite{cafaro2017noxi,muller2015emotion,ringeval2013introducing}, a growing body of work on multi-person interactions has developed in recent years.
Among the social concepts that have been studied in multi-person interactions from a computational perspective are turn-taking \cite{bohus2011multiparty,laskowski2010modeling,ruhlemann2015turn}, laughter~\cite{mckeown2015belfast}, general interest level of the group \cite{gatica2005detecting}, and engagement of individuals inside the group \cite{oertel2013gaze}.
Cohesion is one of the more abstract concepts and describes the tendency of group members to create social bonds and stay united as a group. 
\change{Cohesion is commonly understood as a global measure given by each interactant to the whole group.
In contrast, rapport can be measured for each pair of people within a group and can thus provide a more detailed picture of intra-group relations.}
Hung and Gatica-Perez analysed audio, visual and audio-visual cues to predict cohesion levels in small groups using annotations from external observers~\cite{hung2010estimating}.
More recently, Nanninga et al.\ specifically focused on the connection between group mimicry and task cohesion~\cite{nanninga017cohesion}.
Other works focused on leadership and listener behaviour.
Automatic recognition of emergent leaders is particularly relevant given that those leaders emerge from the interaction among group members (as opposed to designated leaders).
The prediction of perceived leadership alongside dominance, competence and liking in groups of three to four people was also studied using information on speaking activity and speech prosody as well as activity of the body and head~\cite{beyan2017prediction,sanchez2012nonverbal}.
Other works aimed to differentiate
instructed considerate from authoritarian leadership styles~\cite{feese2011discriminating} or, more recently, naturally emerging autocratic or democratic behaviour~\cite{beyan2017prediction}.
Classification of group members into attentive listeners, side participants, and bystanders was studied in \cite{oertel2015deciphering}.

In summary, while a number of works studied different prediction tasks in multi-person interactions, some of which are related to rapport, to the best of our knowledge we are first to predict rapport in a multi-person setting.

\subsection{Rapport}

Among the different concepts of social interactions, rapport is arguably one of the most fundamental and thus important.
Failure to build rapport can result in poor social interactions, decreased collaboration, and worse interpersonal outcomes \cite{burns1984rapport,kelley2014influence,tsui1985failure}.
In an early work, Tickle-Degnen and Rosenthal identified three components that are important for rapport: attention, positivity, and coordination~\cite{tickle1990nature}.
The importance of these components can change over the course of a relationship as can the expression of components of rapport.
For example, insults can help build rapport in later stages of a relationship~\cite{ogan2012rudeness}.
Izard hypothesised connections between personality traits and the ability to build rapport~\cite{izard1990personality}.
For example, people with high extraversion were deemed to find it easier to build rapport given that they might more easily focus their attention on others.
Furthermore, people with a tendency towards negative emotions might not be able to express the positivity component of rapport strongly enough.

Research on the link between dyadic rapport and non-verbal behaviour is extensive, so we only discuss two representative works.
Harrigan, Oxman and Rosenthal analysed rapport ratings for physicians obtained from nurses~\cite{harrigan1985rapport}. 
They found that physicians sitting with uncrossed legs and arms in symmetrical side-by-side positions directly facing the patient received higher rapport ratings.
Bernieri et al.\ conducted an important analysis on dyad rapport and its judgement across different situations~\cite{bernieri1996dyad}.
When comparing subjective rapport ratings with ratings by external observers they found that observers had a hard time rating rapport consistently with the participants who experienced the situation.
Further analyses showed that while observer judgements were mainly based on the amount of expressiveness, self-ratings were adapted to the specific situation at hand.
These results indicate that observer ratings of rapport are not adapted to the specific situation.
We therefore opted to use self-reported rapport ratings in this work.

\subsection{Predicting Rapport in Dyadic Interactions}

Computational approaches to rapport prediction focused on dyadic interactions, typically with the motivation to develop artificial agents that are able to build rapport with users.
The first line of work investigated non-verbal cues for rapport prediction.
For example, Wang and Gratch used selected facial action units (AU) to predict felt rapport in human-human and human-agent interactions~\cite{wang2009rapport}. 
They found that felt rapport was encoded in the absence of AUs encoding negative emotions rather than in the presence of AUs indicating positive emotions.
Hagad et al.\ used participants' postures and their congruences to predict rapport in dyadic interactions~\cite{hagad2011predicting}.
Other works used verbal cues or mixtures of non-verbal and verbal cues for rapport prediction.
A recent study by Cerekovic et al.\ focused on predicting self-reported and observer-rated rapport between humans and virtual agents using verbal and non-verbal cues~\cite{cerekovic2016rapport}.
The results showed that self-reported rapport is rather weakly correlated with observer-judged rapport, and also harder to predict than the latter.
Zhao et al.\ applied temporal pattern mining to extract rules for rapport management in dyads of peer-tutoring strangers and friends~\cite{zhao2014towards,zhao2016socially}.
An example of such a rule indicative of high rapport in friend dyads is the verbal violation of a social norm by one interactant while in parallel her friend is smiling.
Finally, bonding is a concept related to rapport and has recently been studied in the context of dyadic human-human and human-agent interactions \cite{jaques2016understanding}, also depending on personality \cite{jaques2016personality}.

%!TEX root = main.tex
\section{A Dataset of Small-Group Interactions}

Given the lack of suitable datasets for the development and evaluation of algorithms for rapport detection, we designed a human study to collect audio-visual non-verbal behaviour data and rapport ratings during small group interactions. 
Our dataset consists of 22 group discussions in German, each involving either three or four participants and each lasting about 20 minutes, resulting in a total of more than 440 minutes of audio-visual data.

\subsection{Recording Setup}
\begin{figure}
\centering
  \includegraphics[width=1.0\columnwidth]{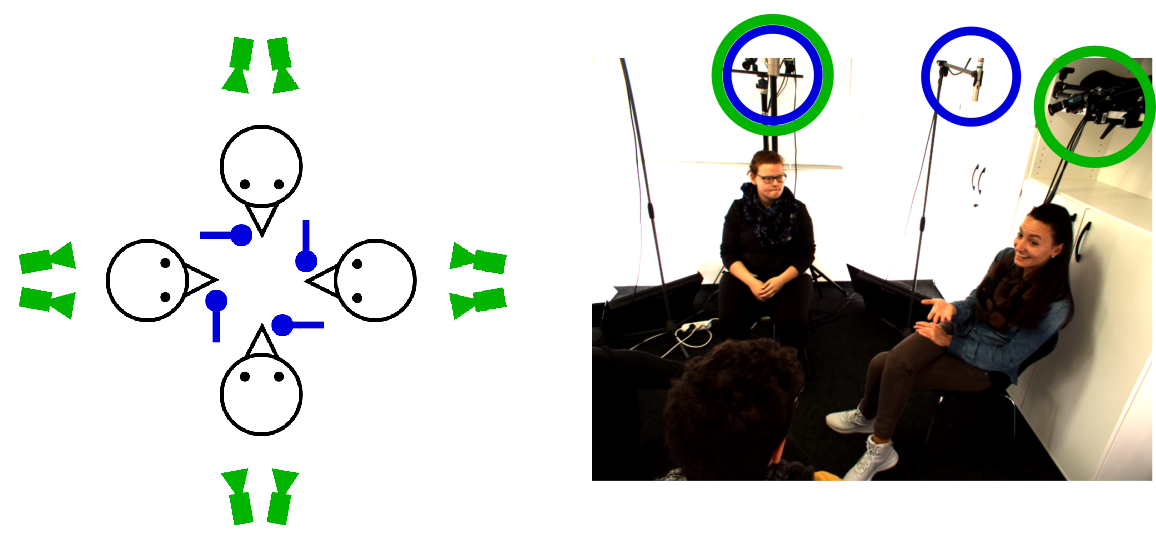}
  \caption{Illustration of camera and microphone positions during a recording session with four participants. Cameras are shown in green, and microphones in blue. Please note all the equipment was placed slightly above the participants to avoid occlusion for video recording. }
  \label{fig:sensor_placement}
\end{figure}

The data recording took place in a quiet office in which a larger area was cleared of existing furniture. The office was not used by anybody else during the recordings.
To capture rich visual information and allow for natural bodily expressions, we used a 4DV camera system to record frame-synchronised video from eight ambient cameras.
Specifically, two cameras were placed behind each participant and with a position slightly higher than the head of the participant (see the green indicators in \autoref{fig:sensor_placement}).
With this configuration a near-frontal view of the face of each participant could be captured throughout the experiment, even if participants turned their head while interacting with each other.
In addition, we used four Behringer B5 microphones with omnidirectional capsules for recording audio.
To record high-quality audio data and avoid occlusion of the faces, we placed the microphones in front of but slightly above participants (see the blue indicators in~\autoref{fig:sensor_placement}).
To synchronise the audio and video streams, we clapped our hands before and after every recording session.

\subsection{Recording Procedure}

We recruited 78 German-speaking participants (43 female, aged between 18 and 38 years) from a German university campus, resulting in 12 group interactions with four participants, and 10 interactions with three participants. 
During the group forming process, we ensured that participants in the same group did not know each other prior to the study.
To prevent learning effects, every participant took part in only one interaction.

Preceding each group interaction, 
we told the participants that first personal encounters could result in various artifacts that we were not interested in.
As a result, we would first do a pilot discussion for them to get to know each other, followed by the actual recording.
We intentionally misled the participant to believe that the recording system would be turned on only \textit{after} the pilot discussion, so that they would behave naturally.
In fact, however, the recording system was running from the beginning and there was no follow-up recording.
To increase engagement, we prepared a list of potential discussion topics and asked each group to choose the topic that was most controversial among group members.
Afterwards, the experimenter left the room 
and came back about 20 minutes later to end the discussion.
Participants were then asked to complete several questionnaires about the other groups members as described below.
Finally, participants were debriefed, in particular about the deceit, and gave free and informed consent to their data being used and published for research purposes.

\subsection{Data Annotation Using Questionnaires}
Although in this work we were only interested in detection of low rapport, with a view to potential other future uses of our dataset, participants were asked to complete three questionnaires about different social aspects relevant for small group interactions. 
All questionnaires were given in German to increase comprehension of the questions and, in turn, obtain more reliable scores.
\begin{itemize}
  \item \textit{Rapport:} Since rapport is a subjective feeling that is hard to gauge through any existing equipment, we followed previous practice using an 18-item-questionnaire~\cite{bernieri1996dyad} to measure rapport from self reports. Responses were recorded on seven point Likert scales. Each participant rated each item for other individuals in the group, yielding two rapport scores for each dyad inside the larger group.
  \item \textit{Leadership, Dominance, Competence, and Liking:} We were also interested in the correlation between rapport and other well-studied aspects in small group interactions. We thus asked participants to complete the questionnaire used in~\cite{sanchez2012nonverbal} that consists of 12 questions about four different sub-scales (leadership, dominance, competence, and liking) which we recorded using seven-point Likert scales.
  \item \textit{Personality:} Finally, each participant also completed the well-established NEO-FFI questionnaire to assess personality traits, including openness to experience, conscientiousness, extraversion, agreeableness, and neuroticism \cite{costa1992revised}.
\end{itemize}

Given that we were mainly interested in the overall degree to which a participant is able to build rapport with others, we aggregated the rapport scores for a target participant by averaging those given to him by the other participants in the group.
Consequently, a low rapport score indicates that a particular participant did not evoke the feeling of rapport in general for the other participants.
We processed the other annotations in the same way (leadership, dominance, competence and liking).

\subsection{Dataset Statistics}

\begin{table}
  \centering
  \begin{tabular}{l | c c }
    & Means
    & Standard Deviations\\
    \midrule
    Rapport & 5.41 & 0.46\\
    Leadership    & 3.71 & 0.94\\
    Dominance     & 4.14 & 0.96\\
    Competence     & 5.22 & 0.87\\
    Liking    & 5.81 & 0.56\\
  \end{tabular}
  \caption{Means and standard deviations of the aggregated annotations obtained from seven-point Likert scales.}~\label{tab:ratings_descr}
\end{table}

\begin{figure}
  \centering
  \includegraphics[width=0.9\columnwidth]{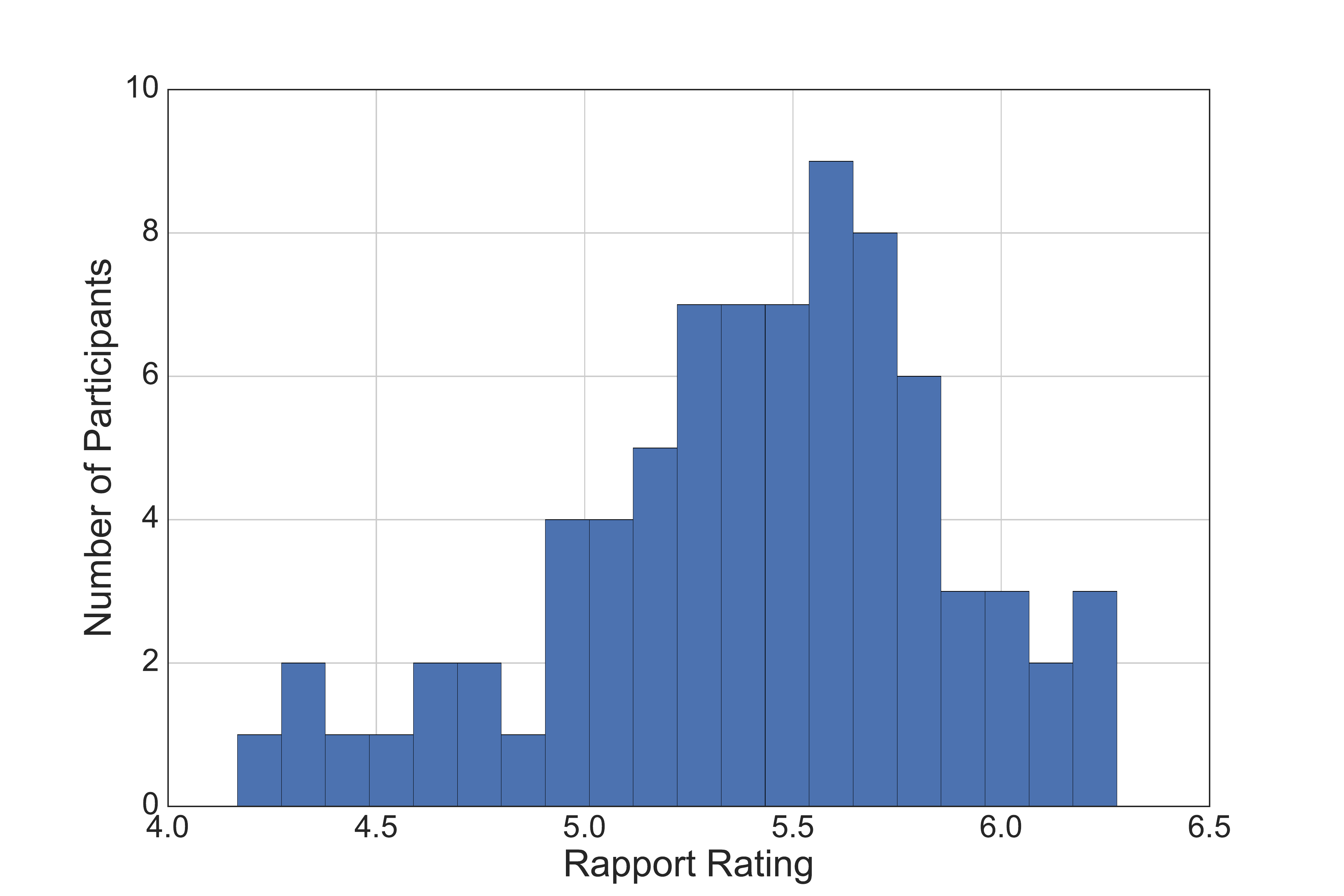}
  \caption{Histogram of the number of participants (y-axis) against the average received rapport ratings from other participants in an interaction (x-axis).}~\label{fig:rapport_distribution}
\end{figure}

Table \ref{tab:ratings_descr} summarises the means and standard deviations of the questionnaire responses over all participants.
Especially for liking, competence and rapport we can observe a tendency towards higher ratings.
A more fine-grained depiction of the distribution of rapport scores is shown in Figure \ref{fig:rapport_distribution}, which shows a tendency towards a left-skewed distribution with a peak at 5.6 and most scores between 5.0 and 6.0. 
The bias towards higher values in questionnaires which involve a potentially more hurtful evaluation of others (liking, competence and rapport in contrast to leadership and dominance) might be due to a general social desirability bias \cite{lavrakas2008encyclopedia}.
Given that we were particularly interested in low rapport,
we grouped the data with the lower 25\% percentile of rapport scores as ``low rapport'' and the rest as ``high rapport''.
This results in 11 interactions without a low-rapport participant (seven of them are three-participant interactions), four interactions with a single low-rapport participant (two three-participant interactions), six interactions with two low-rapport participants (one three-participant interaction), and one interaction with three low-rapport participants (four-participant interaction).

\begin{figure}
  \centering
  \includegraphics[width=0.9\columnwidth]{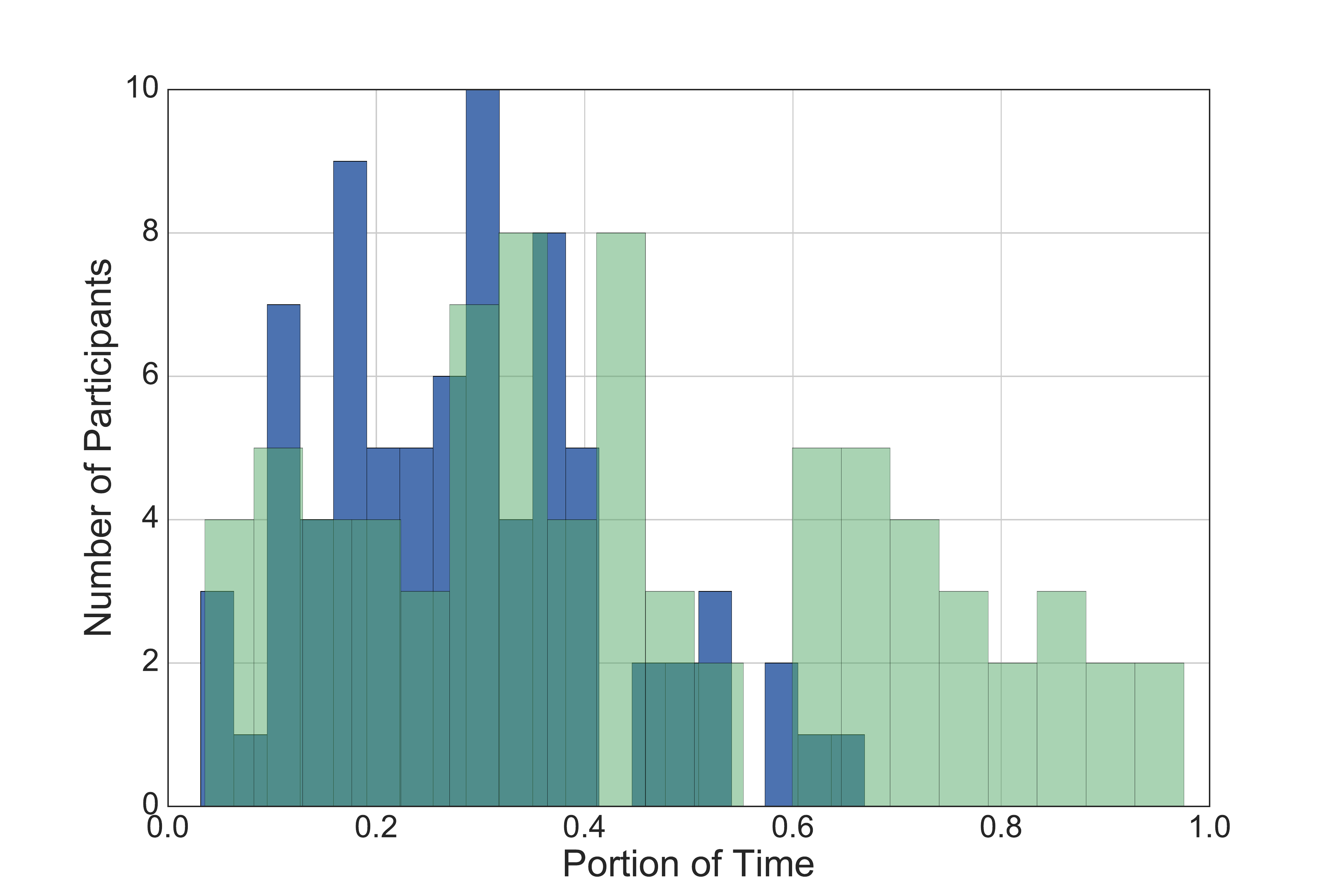}
  \caption{
  Histogram of the number of participants (y-axis) against the portion of time that participants are speaking (blue bars) and smiling (transparent green bars; detected by AU12) during the interactions.
 }~\label{fig:AU_and_speech}
\end{figure}

The diversity of the dataset in terms of participants' behaviour can be illustrated, for example, by the portion of time they spoke and smiled.
Figure \ref{fig:AU_and_speech} shows the histogram of the portion of speaking time (blue bars).
While most participants spoke around 10\% to 40\% of the time per discussion, several participants spoke less than 10\% or more than 50\% of the time.
Moreover, the amount of smiling
is highly diverse across participants (see Figure \ref{fig:AU_and_speech}, transparent green bars).
While some participants hardly smiled at all, others smiled almost constantly.
Table \ref{tab:AU_descr} shows the average activation AUs across participants.
Inspecting the standard deviations, we can see that there is substantial variability in participants' average level of AU activations.

\begin{table*}[!htb]
  \centering
  \begin{tabular}{l | r r r r r r r r r r r r r r r r r r}
    AU& 1&2&4&5&6&7&9&10&12&14&15&17&20&23&25&26&45\\

    \midrule
    $\mu$&0.18 &0.25 &0.36 &0.53 &0.34 &0.43 &0.07 &0.70 &0.44 &0.59 &0.27 &0.39 &0.20 &0.47 &0.20 &0.15 &0.21 \\
$\sigma$&0.08&0.10&0.26&0.25&0.26&0.27&0.06&0.24&0.25&0.26&0.10&0.12&0.10&0.22&0.08&0.07&0.07\\
  \end{tabular}
  \caption{Statistics for average AU activations of all extracted AUs when the participant is not speaking.}~\label{tab:AU_descr}
\end{table*}

%!TEX root = main.tex
\section{Multimodal Method with Non-Verbal Features}

Our multimodal approach to detecting low rapport relies on non-verbal features only, rather than word-related features.
Specifically, it considers facial expression, hand motion, speech activity, and prosodic features. 
In addition, we also exploit synchronisation features and cross-modal features.
The following subsections discuss each of these feature sets.

\subsection{Non-Verbal Features}

\begin{table*}[!htb]
  \centering
  \begin{tabular}{l | l | l}
    Modalities & Notation & Feature Description \\
    \midrule
     Speech   & $TimeSpeak$  & The portion of time the target participant speaks  \\
     Activity & $TimeTurn$ & The average length of speaking turns  \\
              & $RateTurn$ & The number of speaking turns per minute \\
              & $ProbTurn|TurnTrans$ & Probability of taking the turn at turn transition \\
    \midrule
   Prosody    & $PRSx$& Set of 768 prosodic features based on IS09 challenge feature set from \cite{eyben2013recent} \\
    \midrule
      Face    & $PosiFace^{\mu/\sigma}$& Mean and stddev of facial positivity indicator  \\
              & $PosiFace^{\mu/\sigma}_{200s}$& Mean and stddev of facial positivity indicator during the beginning 200 seconds\\
              & $PosiFace^{sync}$ & Amount of synchronisation of facial positivity indicator with other participants \\
              & $Facing$& How much other participants are facing the target participant  \\
              & $MutualFacing$& Amount of mutual facing with other participants  \\
              & $AUx$& Mean intensity of AU$x$  \\
              & $AUx_{200s}$& Mean intensity of AU$x$ during the beginning 200 seconds  \\
              & $AUx^{sync}$& Amount of synchronisation of intensity of AU$x$ with other participants \\
              & $AU^{sync}$& Average amount of synchronisation of all AU intensities \\
              & $ProbAUx$& Probability of AU$x$ being active  \\
              & $ProbAUx_{200s}$& Probability of AU$x$ being active during the beginning 200 seconds\\
              & $ProbAUx^{sync}$& Amount of synchronisation of AU$x$ activation with other participants \\
              & $ProbAU^{sync}$& \makecell[l]{Average amount of synchronisation of all AU activations} \\
    \midrule
     Face and & $AUx_{target|targetSpeak}$& Mean intensity of AU$x$ of target participant when he/she is speaking  \\
     Speech   & $AUx_{target|targetNotSpeak}$& Mean intensity of AU$x$ of target participant when he/she is not speaking  \\
     Activity & $AUx_{other|targetSpeak}$& \makecell[l]{Average mean intensity of AU$x$ of other participants when target\\ participant is speaking}  \\
              & $AUx_{target|otherSpeak}$& \makecell[l]{Average mean intensity of AU$x$ of target participant when another\\ participant is speaking}  \\
              & $ProbAUx_{target|targetSpeak}$ & Probability of AU$x$ of target participant being active when he/she is speaking \\
              & $ProbAUx_{target|targetNotSpeak}$& Probability of AU$x$ of target participant being active when he/she is not speaking \\
              & $ProbAUx_{other|targetSpeak}$& \makecell[l]{Average probability of AU$x$ being active in other participants when target\\ participant is speaking}  \\
              & $ProbAUx_{target|otherSpeak}$& \makecell[l]{Average probability of AU$x$ being active in target participants when another \\participant is speaking} \\
    \midrule
    Hand      & $VelHand$& Average velocity of hands  \\
    Motion    & $VelHand^{sync}$& Amount of synchronisation of hand velocity with other participants  \\
    \midrule
    Hand Motion& $VelHand_{target|targetSpeak}$& Average hand velocity of target participant when he/she is speaking  \\
    and Speech & & \\
    Activity   & & \\
  \end{tabular}
  \caption{Feature notations and descriptions of different modalities. 
  }~\label{tab:features}
\end{table*}

\subsubsection{Speech Activity Features}
Turn-taking is an important attribute in conversations, and there may be a potential link between the turn-taking behaviour in group discussion and felt rapport, for example via reflecting aspects of the coordination component of rapport \cite{tickle1990nature}.
To extract speech activity features, we annotate speaking turns from all recordings.
Based on this information, we compute several features that encode the duration and frequency with which participants speak, and also different characteristics of turn-taking (see Table \ref{tab:features}).

\subsubsection{Prosodic Speech Features}
Apart from the speech activity features, we extract a set of prosodic speech features using openSmile \cite{eyben2013recent}.
We choose the feature set used for the IS09 emotion challenge, as it is a rather small feature set (384 features) and we assume effective features for emotion recognition might also be helpful for rapport detection.
The features are extracted from individual segments when the participant speaks, and then aggregated over all segments of a speaker by taking the mean and the standard deviation, resulting in 768 features.

\subsubsection{Facial Features}
Facial expressions convey informative visual cues of emotions, and they are an important non-verbal channel to express one's feelings and views. 
Therefore, we include facial expression as one of our main features for rapport detection.

Our facial features include head orientations as well as the activation/intensity of facial action units (AUs), and additionally some higher-level facial features built on top of these basic concepts. 
For example, we incorporate features encoding aspects of all three components of rapport suggested by Tickle-Degnen and Rosenthal \cite{tickle1990nature}. 
They include 1) the amount of positivity, 2) interpersonal synchronisation/coordination, and 3) mutual attention reflected by head orientations.
An overview of the facial features is given in Table \ref{tab:features}.

In practice, we used OpenFace \cite{Baltrusaitis2016}.
It is an automatic tool for facial expression analysis that identifies facial landmarks, head pose, and the activation/intensity level of the 17 facial AUs displayed in Table \ref{tab:AU_descr} from a video.
As there are four cameras that cover the face of each participant from different angles, we extract the facial information from all four videos.
Based on the confidence scores given by OpenFace, we selected the best view for each frame and use the facial AUs in this view for further analysis and recognition.
\change{This procedure results in high OpenFace confidence scores (>0.8 on a scale from 0 to 1) in almost all frames (97\%).}

Facial positivity is computed following previous practice~\cite{chikersal2017deep}.
The facial positivity indicator $PosiFace$ for the target participant is set to 1, if AU12 is active, and -1 if AU15 is active in conjunction with at least one of AU1 and AU4 \cite{chikersal2017deep}.
$PosiFace$ is set to 0, when none of the above holds, or when both the positivity (AU12) and negativity AUs (AU1, AU4, AU15) are active.
To reflect the intuition that the first minutes of a discussion are special, as the participants are just getting to know each other, we include additional versions of the above features that only take into account the first 200 seconds of the interaction (e.g.\ $AUx_{200s}$).
Face orientation features ($Facing$ and $MutualFacing$) are constructed by thresholding of the face orientation estimated from the frontal view of the target participant.
Additionally, we extract various features to describe the synchronisation of facial expressions among participants.
The general approach to computing synchronisation of features between participants is detailed below.

\subsubsection{Synchronisation Features}

Inspired by the findings that 1) mirroring is an important phenomenon that can reflect rapport and facilitate the building of rapport \cite{bernieri1988coordinated} and that 2) synchronisation/coordination is one of three basic components of rapport \cite{tickle1990nature}, we build features to delineate the amount of behavioural synchronisation between participants.
To measure the feature synchronisation of two participants, we compute the distance between the pair of feature signals using Dynamic Time Warping (DTW) with a Sakoe-Chiba band of five seconds \cite{sakoe1978dynamic,chikersal2017deep}.
We then compute the amount of synchronisation of a target participant $i$ with all other participants in the interaction, by averaging the DTW distances of the target participant to others.
In other words, for a feature signal $F_i$, the resulting average synchronisation is 
$\sum_{j\in N\setminus\{i\}} $DTW$(F_i,F_j)$, 
where $N$ is the set of all participants in the interaction where the target participant $i$ takes part, and DTW denotes the DTW function.

\subsubsection{Hand Motion Features}
Body posture and its coordination among people can be indicative of rapport \cite{bernieri1988coordinated}.
Since in our study setup all the participants were sitting, we focuse on hand motion.
We use the multi-person pose estimation method OpenPose \cite{cao2017realtime} to extract poses from videos.
OpenPose extracts the joint locations of the human body from the 2D video data.
\change{Based on the frames in which both hands are detected (on average 77\%)}, we compute several features, such as the total amount of hand movement for each participant as well as the synchronisation of hand movements between participants (see Table \ref{tab:features} for details).

\subsubsection{Cross-Modal Features}

In addition to the unimodal features described above, prior research pointed out that the coordination between different modalities, such as gaze-hand coordination, can reflect human mental states~\cite{huang2016stressclick}.
Moreover, cross-modal features have been applied in the context of leadership prediction \cite{beyan2017prediction}.

We design a number of cross-modal features, specifically to encode participants' evaluation of each other by analysing their facial expressions while others are speaking, and also to compensate for the influence on AU detection during speaking.
These features include 1) AU activations and intensities while the participant is speaking or not speaking, 2) average AU activations and intensities of all other participants while the target participant is speaking, and 3) AU activations and intensities of the target participant while other participants are speaking.
Apart from AUs, we combine hand motion information with speech activity into a feature that measures the amount of hand movement while the participant is speaking.
This feature is intended to encode how much a participant is gesticulating during speaking.

\subsection{Learning Low Rapport Using an Ensemble of SVMs}

We train Support Vector Machines (SVM) with radial basis function kernels to classify participants' received rapport ratings into low versus medium-to-high rapport. 
The cost parameter $C$ of SVM is tuned in a nested inner validation loop.
We use a leave-one-interaction-out cross-validation scheme to evaluate the performance of our models.
As a performance metric, we choose average precision (AP), which is common for detection problems, as it is better suited to measure the performance of models on data with class imbalances than, for example, accuracy.
The necessary ranking of test examples is obtained by using probability estimates of the SVM.
To marginalise out any fluctuations due to random initializations of the SVM optimisation method, we train 1,000 SVMs, from which we extract ensemble predictions by averaging.

%!TEX root = main.tex

\section{Experimental Results}

In the experimental evaluation of our proposed approach to low rapport detection, we quantify the contribution of different feature sets and individual features, as well as the amount of information that can be exploited from different temporal segments of an interaction.
Finally, we show additional results concerning the relation of rapport to other previously investigated concepts in small groups.

\subsection{Identifying Important Feature Sets}

\begin{figure*}
  \centering
  \includegraphics[width=\textwidth]{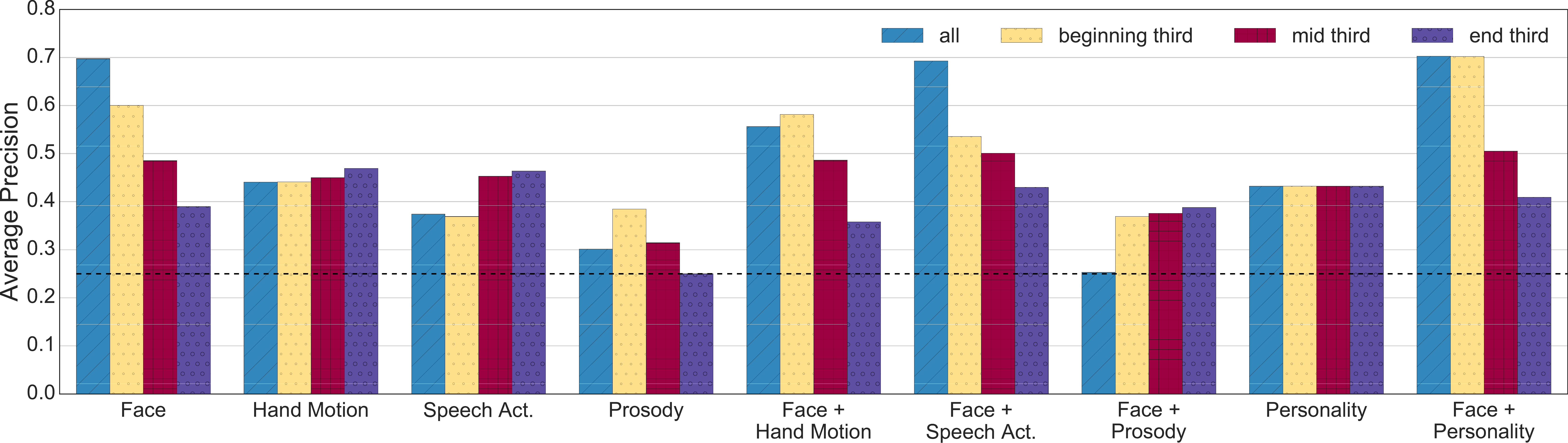}
  \caption{Performance of different feature sets (groups along x-axis) across temporal segments for feature extraction (colour). From left to right, the first four groups indicate the performances using unimodal feature sets, followed by three groups of performances using two modalities, and another two using personality and with facial features. The dotted line indicates the performance of a random predictor.
  }~\label{fig:feature_comparison}
\end{figure*}

To understand the contribution of different modalities to recognition of low rapport, we evaluate our approach with different subsets of features.
Figure \ref{fig:feature_comparison} shows the performance comparison.
The x-axis presents different feature sets. 
Bars with different colour represent the performances of models using different temporal segments for feature extraction, i.e. from the whole interaction (blue), and the first (yellow), middle (red), and last (purple) third of the interaction. 
Since we define the 25 percentile of our data with the lowest score as low rapport, the baseline method (dashed line) that ranks the test data randomly results in 0.25 AP.

In this subsection, we focus on results on full interaction data only (blue bars), for which the overall highest performance is achieved by facial features (0.7 AP).
They perform significantly better than the other unimodal feature sets (see the first four groups in Figure \ref{fig:feature_comparison}).
However, hand motion, speech activity and prosodic feature sets can also outperform the baseline (0.37, 0.44, 0.30 AP, respectively), indicating that each modality carries a certain amount of useful information for low rapport detection.

Surprisingly, adding additional features to the facial features does not further improve the performance for whole-interaction data.
Specifically, adding speech activity and cross-modal combinations of speech activity and facial features (Face + Speech Act.) achieves a comparable result (0.69) to Face alone (0.7).
The combination of face and hand motion features (Face + Hand Motion) produces an AP of 0.56, whereas combining face and speech activity (Face + Speech Act.)\ or prosodic features (Face + Prosody) yields a baseline performance.
All possible further combinations of feature sets fail to improve performance.
This result implies that facial features play an important role for rapport detection in group interactions.

\begin{figure}
  \centering
  \includegraphics[width=\columnwidth]{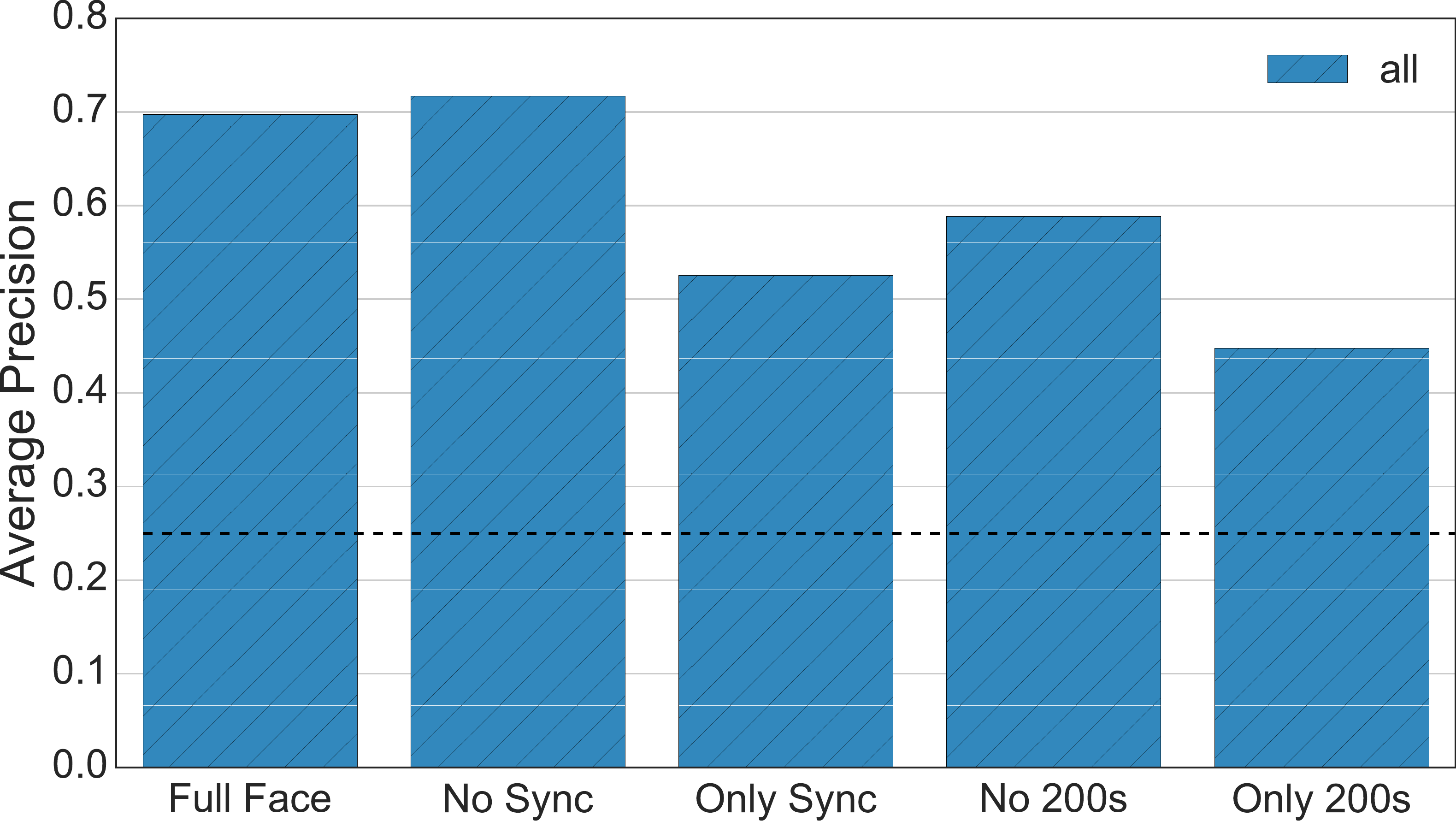}
  \caption{Results of ablation studies on facial feature set. From left to right: full set of facial features, without synchronisation features, only synchronisation features, without features extracted from the beginning 200s, only features extracted from the beginning 200s.}~\label{fig:face_ablation}
\end{figure}

To further understand which types of facial features lead to good performance, we perform an ablation analysis (see Figure \ref{fig:face_ablation}).
Firstly, we split face features into four groups: 1) synchronisation features,
2) non-synchronisation features, 3) without using facial features extracted in the beginning 200s of each interaction, and 4) using only those features that were extracted in the first 200s of an interaction.
Surprisingly, it turns out that facial features without synchronisation even outperform Face (comprising both sync and non-sync features), though with a marginal improvement (0.72 AP).
In contrast, facial features with synchronisation only result in 0.53 AP.
Thus, although facial synchronisation features carry a certain amount of information about rapport, the mirroring and behavioural coordination effects encoded in them are not indicative enough to improve over the basic facial features.
Still, it could be possible that mirroring of particular member(s), or at particular points in time in the interaction (e.g.\ while speaking), may have a stronger indication of rapport, which needs further investigation.
We also see that including the features extracted from the beginning 200s contribute to an improvement (Face vs. No 200s), though using these features alone has a low AP (0.45).
This indicates that features extracted at the beginning of an interaction have a special relation to rapport, complementary to features extracted from the full interaction. 

Finally, we study how well low rapport can be predicted from personality scores, a factor that can be measured without observing the actual interaction.
To this end, we train a SVM on NEO-FFI scores of the target participants, which leads to an AP of 0.43.
Although training on personality scores alone does not give a high rapport recognition performance, it can clearly outperform the baseline (0.25), and even speech activity (0.37) and prosodic features (0.30), and yields a comparable result with that of hand motion (0.44) from the actual interaction.
This finding is interesting, as it indicates that if an intelligent user interface can gather information on personal traits, e.g. from a personal device, there is a high chance that it can make a correct prediction of rapport in a future group interaction even without access to the actual interaction signals.

\subsection{Prediction from Temporal Segments}
In addition to understanding the contribution of different feature sets, we also evaluate the amount of information that our method is able to exploit from different temporal segments of the interactions.
Specifically, we divide each interaction into three segments and train and test on each segment of the interactions only.
Figure \ref{fig:feature_comparison} shows the average precisions achieved in these three cases (yellow, red, and purple bars).

For our best-performing feature set on full recordings, facial features without synchronisation features, we can observe a clear trend that the amount of useful information diminishes over the time of the interaction.
This indicates that rapport is encoded in facial behaviour especially at the beginning of an interaction.
However, other parts of the interaction carry complementary information, as the performance of facial features for the first third (0.60 AP) is significantly lower compared to the corresponding performance for the whole interaction (0.70 AP).
Moreover, it is very encouraging to see that if personality information is available, facial features extracted only from the first third of the interaction can successfully reach the best performance that can be achieved using the entire interaction.
This result implies a promising application scenario, since it indicates a prior personality measurement can help to make accurate predictions with only short observations of additional behaviour.
This potentially allows for effective interventions to support group interactions at an early stage.

\begin{table}[t]
  \centering
  \begin{tabular}{l | r }
    Feature& $t$-score\\
    \midrule

    $AU09$ &  2.19\\
    $AU23$ &  2.04\\
    $AU02_{before200}$ & 1.78\\
    $MutualFacing$ &  1.77\\
    $ProbAU25$ & 1.75\\
    $AU25$ & 1.50\\
    $AU14$ & 1.41\\
    \midrule
    $PosiFace^{\sigma}$           & -1.93 \\
  \end{tabular}
  \caption{Features from the face feature set with the highest absolute t-scores for discriminating between low and high rapport.}~\label{tab:feature_scores}
\end{table}

\subsection{Identifying Important Features}
This section extends the previous evaluation to a finer granularity, by investigating the contribution of individual features on the best-performing feature set (facial features without synchronisation features).
The presented results are obtained from whole interaction data.
To identify how well individual features can discriminate between low and high rapport we compute t-scores for each feature separately.
\change{T-scores measure the linear dependency between features and the target (in our case: low vs.\ medium-to-high rapport).
The higher the absolute value of a t-score, the more likely it is that a linear dependence exists in the population.
In addition, the sign of the t-score indicates the direction of the dependency, making them straightforward to interpret.
It is important to note that our trained SVMs might also use nonlinear dependencies in the data, which cannot be reflected in t-scores.}
A list of features with the highest absolute t-scores is given in Table \ref{tab:feature_scores}.

According to these results, low rapport is especially associated with the average intensities of AU9 (nose wrinkler), AU23 (lip tightener), AU2 (outer brow raiser) during the beginning 200s, AU25 (lips part) and AU14 (dimpler), as well as the probability of AU25 being active.
AU9 is often seen in disgust or anger, AU23 in sadness, and AU2 in surprise, fear, disgust or anger \cite{ghayoumi2016unifying}.
This is in line with prior work finding that low rapport is encoded in the presence of facial AUs associated with negative emotions \cite{wang2009rapport}.
AU25 on the other hand indicates speaking, meaning that a large amount of talking is indicative of low rapport.
This is confirmed by the strong dependency between the amount of speaking and low rapport (t=2.9).
A bit surprisingly, AU14 seems to be indicative of low rapport although this AU is often present in facial displays of happiness~\cite{ghayoumi2016unifying}.
\change{Moreover, our results show that a lot of mutual facing is indicative of low rapport.
As mutual facing can be seen as a proxy for attention, this result seems to contradict the theory put forward by Tickle-Degnen and Rosenthal~\cite{tickle1990nature} who postulated that a high degree of mutual attention is indicative of high rapport.
The most likely reason for the negative connection observed in our interaction context is that mutual facing is more frequent in participants who speak a lot, resulting from the social convention of facing the current speaker.
A lot of speaking, in turn, seems to be related to low rapport.
As such, this finding underlines the strong context dependency of the connection between nonverbal behaviour and rapport.}

\subsection{Understanding Correlations Among Group Attributes}

\begin{table}
  \centering
  \begin{tabular}{l | r r r r r}
    & Lead 
    & Dom  
    & Com  
    & Like 
    & Rap\\
    \midrule
    Lead &  & \textbf{0.80} & \textbf{0.41} & 0.01 & \textbf{0.39} \\
    Dom  &  &      & \textbf{0.50} & 0.08 & \textbf{0.52} \\
    Com &  &      &      & \textbf{0.31} & \textbf{0.70} \\
    Like     &  &      &      &      & \textbf{0.52} \\
    \midrule
    O     & 0.01  & 0.10  & 0.21  & 0.02  & 0.15  \\
    C     & -0.09 & -0.13 & -0.06 & -0.04 & -0.13 \\
    E     & 0.12  & 0.12  & -0.00 & 0.17  & 0.16  \\
    A     & -0.22 & -0.11 & -0.07 & \textbf{0.30}  & 0.04  \\
    N     & \textbf{-0.25} & \textbf{-0.32} & -0.18 & 0.10  & -0.21 \\
  \end{tabular}
  \caption{Pearson correlations coefficients between interaction attributes. The lower part of the Table shows correlations between personality scores and the rest interaction attributes. Bold coefficients indicate statistical significance at $\alpha=0.05$, two-tailed.
  }~\label{tab:questionnaireCorrs}
\end{table}

As we are the first to propose the detection of low rapport in a multiparty conversation setting where all participants rate each other, it is important to investigate how this concept of rapport is related to the existing concepts that have been studied in multiparty conversations in the literature.

In particular, the attributes (leadership, dominance, competence, and liking) proposed by Sanchez-Cortes et al.~\cite{sanchez2012nonverbal} are relevant to our work and are measured via the same paradigm as ours. 
That is, every participant rates every other participant within the group.
Moreover, especially the PLike scale suggested in their work~\cite{sanchez2012nonverbal} seems conceptually close to rapport.
In contrast, it is difficult to directly compare with cohesion, as it is a group-level attribute~\cite{chin1999perceived}.
To investigate the association between rapport and other group interaction attributes, we compute the aggregated score in the same way as we process rapport.
Specifically, we average all ratings a participant received from other participants. 
We then calculate the Pearson correlation coefficient between the resulting scores.

Table \ref{tab:questionnaireCorrs} gives the correlations between different interaction attributes.
It is interesting to see that rapport shows a strong correlation with competence (0.70), an obvious correlation with dominance (0.52) and liking (0.52), and a moderate correlation with leadership (0.39).
The correlation analysis also reveals that although rapport is highly associated with competence, they are rather different with respect to their correlation with liking (rapport: 0.52; competence: 0.31).
This implies that rapport is a complex construct associated with multiple different interaction attributes.

\change{Finally,
we computed Pearson correlation coefficients between personality scores and rapport (see Table~\ref{tab:questionnaireCorrs}).
Although not significant in a two-tailed test, the small negative correlation of rapport with neuroticism and the small positive correlation with extraversion is in line with hypotheses on the connection between rapport and personality found in prior work~\cite{izard1990personality}.
As with our previous feature analysis, it is important to keep in mind that the SVM might exploit nonlinear dependencies which are not reflected in the correlations.}

In general, the correlations between rapport and different interaction attributes corroborate our hypothesis that rapport is a concept pertinent to but considerably distinct from the existing attributes proposed in previous studies~\cite{costa1992revised,sanchez2012nonverbal}.

%!TEX root = main.tex

\section{Discussion}
\iffalse
\begin{itemize}
    \item best performance first third of interaction is good for applications that want to make intervention quickly. having data from full conversation is still better however
    \item high ranking of AU06 based features in contrast to prior study in human-avatar interaction setting
    \item prediction using personality only is a good baseline to see that observing actual novnerbal behavior is important (of course, one could also have other questionnaires closer to rapport...)
    \item set theory of personality from tickle degnen commentary in relation to observed correlations of personality with rapport
\end{itemize}
\fi

In this work we proposed a multimodal approach for detecting low rapport in small group interactions.
To the best of our knowledge, we are the first to conduct such an investigation, taking into consideration individual behavioural features from separate modalities (e.g. facial expression and speech activity), cross-modal features (e.g.\ hand motion while speaking), as well as high-level interaction signals (e.g.\ behavioural mirroring).
Evaluations on a novel 78-participant dataset, the first of its kind, showed that facial expressions are, in general, the most powerful signal for low rapport detection.
We further demonstrated that incorporating participants' personality into our pipeline could improve performance for early prediction. 
This is encouraging, as recent years have seen an increase of methods to automatically predict personality traits of an individual user~\cite{vinciarelli2014survey}, e.g.\ using mobile phones~\cite{de2013predicting} or eye movement analysis~\cite{Hoppe15_ubicomp}.
These methods could help improve early rapport prediction without requiring additional explicit user input in the from of personality questionnaires.

The possibility to predict low rapport early and accurately enables next-generation ambient intelligent systems with the ability to support users if they fail to establish rapport with each other.
\change{Such systems could, for example, use ambient displays to encourage or amplify behaviour known to improve rapport~\cite{balaam2011enhancing}.
Advice for the whole group could involve proposing different interaction strategies or even socialising games to increase rapport, or encourage other people to take over or lead the discussion.
Individual advice could be provided on personal screens or head-mounted displays~\cite{damian2015augmenting,schiavo2014overt}.}
Beyond the small group setting, we believe automatic detection of low rapport also has potential for applications in autism spectrum disorders, e.g.\ by supporting people with this disorder in properly interpreting rapport in interactions with others or even helping them to notice low rapport at all.
To execute effective support strategies in these settings, it will be particularly important to detect low rapport at an early stage of the interaction.
Encouragingly, our approach is able to achieve this goal when incorporating prior knowledge of personality.
In addition, our results showed that facial features alone can achieve high performance given information on the entire interaction.
As cameras and microphones become pervasive in personal devices, low rapport detection could become a key component in many intelligent user interfaces that aim to positively influence daily social interactions, reduce stress, avoid conflicts, and thus lead to harmonious computer-mediated interactions.

Our results also suggest that a prediction performance above chance can still be reached if certain modalities are unavailable.
This implies the ability of low rapport detection to adapt to diverse interaction settings.
In practice, our method therefore can suggest an alternative modality combination in case the best modality is temporarily inaccessible.
Even when there is no data from the actual interaction at all, an educated guess can be made based on the prior knowledge of personality scores in order to support those who are most likely to fail in establishing rapport with others.
Given all this, our method has significant potential to pave the way for rapport-aware computer-meditated communication.

Despite these promising results, there are some limitations that we plan to address in future work.
Our results showed that facial expressions are the most indicative modality.
However, analysing multimodal signals using a more sophisticated model, such as a neural network, might allow use of information from multiple modalities more efficiently and achieve an even higher recognition accuracy. 
Furthermore, the present study was conducted in a controlled laboratory environment.
While this is in line with prior works on rapport during dyadic interactions and beneficial for experimental control, it will be interesting to investigate how our findings can generalise to in-the-wild situations, e.g.\ interactions at home, and combining the analysis of rapport with other personal or social signals that can be captured using mobile devices or on-body sensors.

%!TEX root = main.tex

\section{Conclusion}

This work proposed the first audio-visual multimodal approach to low rapport detection in small group interactions.
We evaluated our method on a novel 78-participant dataset consisting of 22 three- and four- person discussions.
We studied a diverse set of non-verbal behaviours, including facial expressions and orientations, hand motion, speech activities, and prosodic features as well as higher-level interaction signals, e.g.\ reflecting mirroring effects. 
Our results showed that facial features in general are most indicative to detect failure in establishing rapport in group interactions.
Moreover, adding personality traits allows us to predict low rapport early on in the interaction.
As such, our study advances the understanding of non-verbal behaviour and rapport establishment, pointing the way towards new intelligent user interfaces that incorporate low rapport detection to prevent disharmony in social interactions on the~fly.

\section{Acknowledgments}
This work was supported, in part, by the Cluster of Excellence on Multimodal Computing and Interaction at Saarland University, Germany, as well as a JST CREST research grant (JPMJCR14E1), Japan.

%\balance{}
% BALANCE COLUMNS
\balance{}

% REFERENCES FORMAT
% References must be the same font size as other body text.
\bibliographystyle{SIGCHI-Reference-Format}
\bibliography{references}

\end{document}